%% file: main.tex
\documentclass[letterpaper,twocolumn,10pt]{article}
\usepackage{usenix,epsfig,endnotes}

\usepackage{amsmath,amssymb,amsfonts}
\usepackage{graphicx}

\usepackage{textcomp}
\usepackage{xcolor}
\usepackage{hyperref}
\usepackage{listings}
\usepackage{booktabs}
\usepackage{multirow}
\usepackage{url}
\usepackage{xspace}
\usepackage{enumitem}
\usepackage{tikz}
\usetikzlibrary{shapes.geometric, arrows.meta, positioning, fit, backgrounds, calc, decorations.pathreplacing}

\usepackage[most]{tcolorbox}

\newcommand{\system}{\textsc{PIDSMaker}\xspace}

\definecolor{codebackground}{RGB}{246, 248, 250}
\definecolor{codeborder}{RGB}{208, 215, 222}

\definecolor{bashtitle}{RGB}{36, 41, 47}
\definecolor{bashkeyword}{RGB}{207, 34, 46}
\definecolor{bashstring}{RGB}{10, 48, 105}
\definecolor{bashcomment}{RGB}{110, 119, 129}
\definecolor{bashprompt}{RGB}{87, 96, 106}

\definecolor{yamltitle}{RGB}{88, 110, 117}
\definecolor{yamlkey}{RGB}{51, 51, 51}
\definecolor{yamlvalue}{RGB}{102, 102, 102}
\definecolor{yamlcomment}{RGB}{170, 170, 170}

\definecolor{pythontitle}{RGB}{76, 86, 106}
\definecolor{pythonkeyword}{RGB}{51, 51, 51}
\definecolor{pythonstring}{RGB}{102, 102, 102}
\definecolor{pythoncomment}{RGB}{170, 170, 170}
\definecolor{pythonbuiltin}{RGB}{80, 80, 80}

\lstdefinestyle{basestyle}{
    basicstyle=\ttfamily\small,
    breaklines=true,
    breakatwhitespace=false,
    showstringspaces=false,
    tabsize=2,
    xleftmargin=0pt,
    xrightmargin=0pt,
    aboveskip=0pt,
    belowskip=0pt,
}

\lstdefinestyle{bashstyle}{
    style=basestyle,
    language={},                              
    basicstyle=\ttfamily\small\color{black},
    commentstyle=\color{bashcomment}\itshape,
    morecomment=[l]{\#},
    emph={./run.sh},                          
    emphstyle=\color{bashhighlight}\bfseries,
}

\newtcblisting{bashcode}[1][]{
    enhanced,
    breakable,
    listing only,
    listing style=bashstyle,
    colback=codebackground,
    colframe=codeborder,
    boxrule=0.8pt,
    arc=4pt,
    left=8pt,
    right=8pt,
    top=12pt,
    bottom=6pt,
    toptitle=2pt,
    bottomtitle=2pt,
    fonttitle=\small\bfseries\sffamily,
    coltitle=white,
    colbacktitle=bashtitle,
    attach boxed title to top left={xshift=10pt, yshift=-\tcboxedtitleheight/2},
    boxed title style={
        colback=bashtitle,
        colframe=bashtitle!80!black,
        boxrule=0.5pt,
        arc=3pt,
        left=6pt,
        right=6pt,
        top=0pt,
        bottom=0pt,
    },
    overlay={
        \node[anchor=north east, inner sep=4pt, font=\footnotesize\ttfamily, text=bashprompt]
            at (frame.north east) {\$};
    },
    #1
}

\lstdefinelanguage{yaml}{
    keywords={true,false,null,y,n},
    sensitive=false,
    comment=[l]{\#},
    morestring=[b]',
    morestring=[b]",
}

\definecolor{yamlpipeline}{RGB}{0,100,150}  

\lstdefinestyle{yamlstyle}{
    style=basestyle,
    language=yaml,
    basicstyle=\ttfamily\small\color{black},
    keywordstyle=\color{yamlvalue}\bfseries,
    keywordstyle=[2]\color{yamlpipeline}\bfseries,  
    morekeywords=[2]{construction, transformation, featurization, batching, training, evaluation, triage},  
    commentstyle=\color{yamlcomment}\itshape,
    stringstyle=\color{yamlvalue},
    moredelim=[l][\color{yamlkey}\bfseries]{:},
    moredelim=**[s][\color{yamlkey}]{-\ }{:},
    literate={:}{{\textcolor{black}{:}}}1,
}

\newtcblisting{yamlcode}[1][]{
    enhanced,
    breakable,
    listing only,
    listing style=yamlstyle,
    colback=codebackground,
    colframe=codeborder,
    boxrule=0.8pt,
    arc=4pt,
    left=8pt,
    right=8pt,
    top=12pt,
    bottom=6pt,
    toptitle=2pt,
    bottomtitle=2pt,
    fonttitle=\small\bfseries\sffamily,
    coltitle=white,
    colbacktitle=yamltitle,
    attach boxed title to top left={xshift=10pt, yshift=-\tcboxedtitleheight/2},
    boxed title style={
        colback=yamltitle,
        colframe=yamltitle!80!black,
        boxrule=0.5pt,
        arc=3pt,
        left=6pt,
        right=6pt,
        top=0pt,
        bottom=0pt,
    },
    #1
}

\lstdefinestyle{pythonstyle}{
    style=basestyle,
    language=Python,
    basicstyle=\ttfamily\small\color{black},
    keywordstyle=\color{pythonkeyword}\bfseries,
    stringstyle=\color{pythonstring},
    commentstyle=\color{pythoncomment}\itshape,
    emphstyle=\color{pythonbuiltin},
    emph={self, True, False, None},
}

\newtcblisting{pycode}[1][]{
    enhanced,
    breakable,
    listing only,
    listing style=pythonstyle,
    colback=codebackground,
    colframe=codeborder,
    boxrule=0.8pt,
    arc=4pt,
    left=8pt,
    right=8pt,
    top=12pt,
    bottom=6pt,
    toptitle=2pt,
    bottomtitle=2pt,
    fonttitle=\small\bfseries\sffamily,
    coltitle=white,
    colbacktitle=pythontitle,
    attach boxed title to top left={xshift=10pt, yshift=-\tcboxedtitleheight/2},
    boxed title style={
        colback=pythontitle,
        colframe=pythontitle!80!black,
        boxrule=0.5pt,
        arc=3pt,
        left=6pt,
        right=6pt,
        top=0pt,
        bottom=0pt,
    },
    #1
}

\newcommand{\eg}{e.g.,\@\xspace}
\newcommand{\etal}{et al.\@\xspace}
\newcommand{\vs}{vs.\@\xspace}

\newcommand{\noindgras}[1]{\noindent{\bf #1}}

\def\Snospace~{\S{}}

\begin{document}

\date{}

\title{\Large \bf \raisebox{-0.2\height}{\includegraphics[height=1.5em]{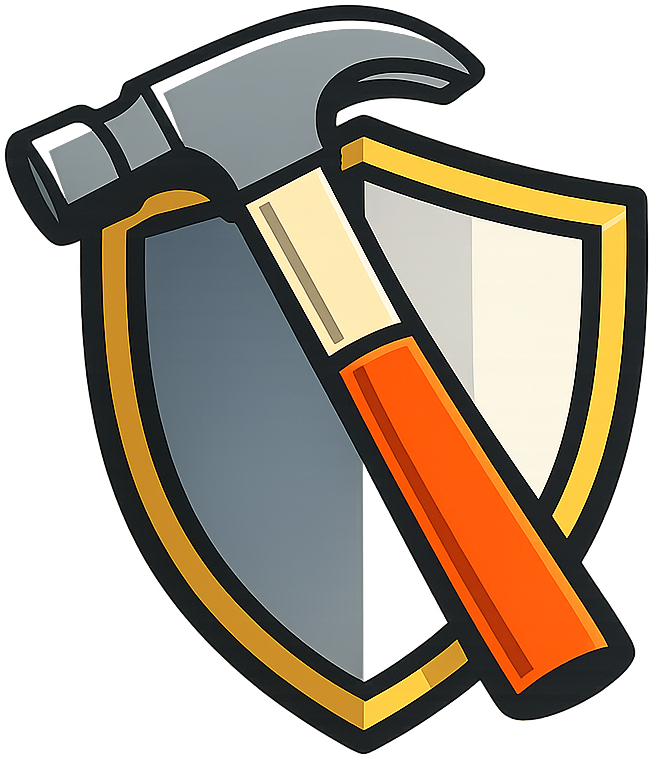}} \system: Building and Evaluating Provenance-based \\Intrusion Detection Systems}


\author{
{\rm Tristan Bilot}\\
University of British Columbia
\and
{\rm Baoxiang Jiang}\\
Xi’an Jiaotong University
\and
{\rm Thomas Pasquier}\\
University of British Columbia
}


\maketitle

\thispagestyle{empty}

\begin{abstract}
\input{sections/abstract}
\end{abstract}

\section{Introduction}
\label{sec:introduction}
\input{sections/introduction}

\input{fig/archi}

\section{Background}
\label{sec:background}
\input{sections/background}

\section{Framework Design}
\label{sec:design}
\input{sections/design}

\section{Use Cases}
\label{sec:usecases}
\input{sections/usecase}

\section{Related Work}
\label{sec:related}
\input{sections/rw}

\begin{figure}[t]
	\centering
	\includegraphics[width=\columnwidth]{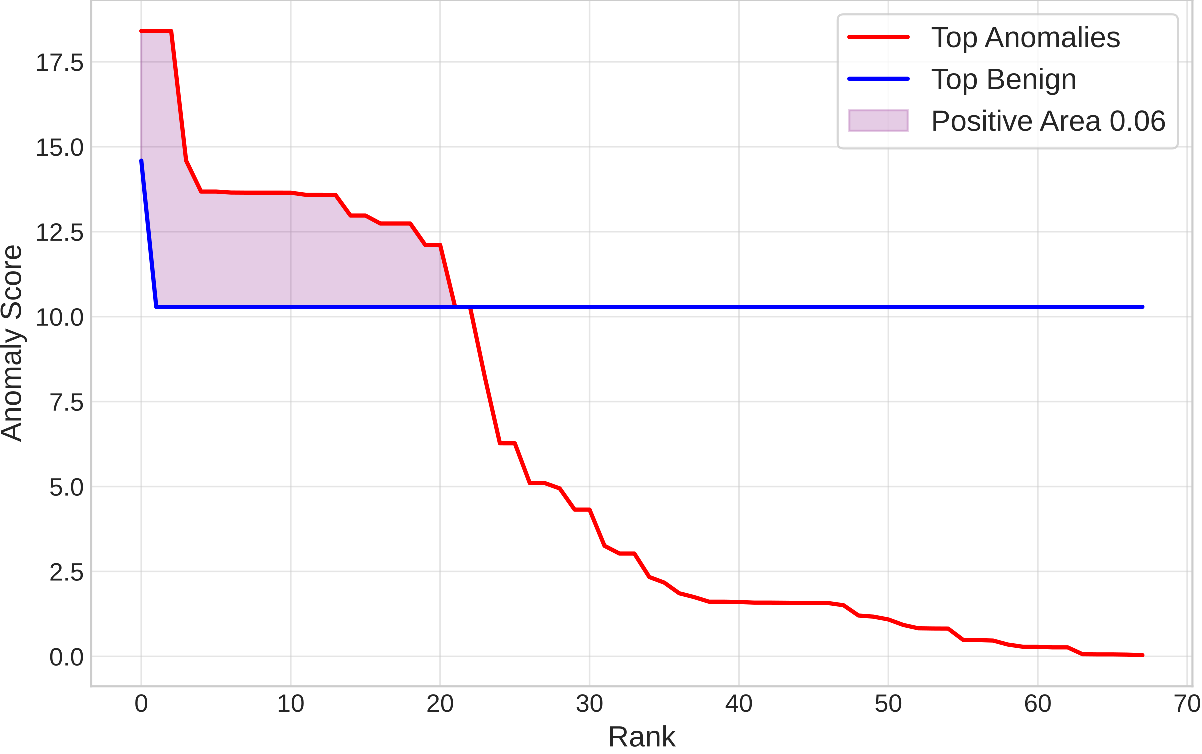}
	\caption{Anomaly scores predicted for top-ranked nodes (\texttt{E3-CADETS} dataset).}
	\label{fig:rank}
\end{figure}

\section{Discussion}
\label{sec:discussion}
\input{sections/discussion}

\section{Community Adoption and Contributions}
\label{sec:community}
\input{sections/community}

\section{Conclusion}
\label{sec:conclusion}
\input{sections/conclusion}

\section*{Availability}
\input{sections/availability}

\section*{Ethics Considerations}
\input{sections/ethics}

\bibliographystyle{acm}
\bibliography{biblio}
\end{document}

%% file: sections/abstract.tex
Recent provenance-based intrusion detection systems (PIDSs) have demonstrated strong potential for detecting advanced persistent threats (APTs) by applying machine learning to system provenance graphs.
However, evaluating and comparing PIDSs remains difficult: prior work uses inconsistent preprocessing pipelines, non-standard dataset splits, and incompatible ground-truth labeling and metrics. 
These discrepancies undermine reproducibility, impede fair comparison, and impose substantial re-implementation overhead on researchers.
We present \system~\footnote{This paper presents \system \href{https://github.com/ubc-provenance/PIDSMaker/releases/tag/2.0.0}{version 2.0.0}.}, an open-source framework for developing and evaluating PIDSs under consistent protocols. 
\system consolidates eight state-of-the-art systems into a modular, extensible architecture with standardized preprocessing and ground-truth labels, enabling consistent experiments and apples-to-apples comparisons. 
A YAML-based configuration interface supports rapid prototyping by composing components across systems without code changes.
\system also includes utilities for ablation studies, hyperparameter tuning, multi-run instability measurement, and visualization, addressing methodological gaps identified in prior work. 
We demonstrate \system through concrete use cases and release it with preprocessed datasets and labels to support shared evaluation for the PIDS community.

\noindent \textbf{Code}: {\small\url{https://github.com/ubc-provenance/PIDSMaker}}\\
\noindent \textbf{Docs}: {\small\url{https://ubc-provenance.github.io/PIDSMaker}}

%% file: sections/introduction.tex
Provenance-based security techniques have gained substantial traction in the security community over the past decade~\cite{unicorn:han:2020,Wang2020YouAW,Han2020SIGLSS,Wang2021THREATRACEDA,Li2023NODLINKAO,Jia2023MAGICDA,Cheng2023KairosPI,wang2024incorporating,Zengy2022SHADEWATCHERRC,Yang2023PROGRAPHERAA,Rehman2024FlashAC,Goyal2024RCAIDER,King2022EulerDN,Hossain2018SLEUTHRA,Xie2016UnifyingID,Manzoor2016FastMA,Han2017FRAPpuccinoFT,Lemay2017AutomatedPA,Xie2020PagodaAH,Kapoor2021PROVGEMAP,King2023EdgeTorrentRT,Meng2024PGAIDAA,Xu2024ProcSAGEAE,inam2023sok,bilot2023graph,Ayoade2020EvolvingAP,jian2025,bilot2025simpler}. 
These techniques capture system execution as directed graphs: nodes denote system entities (e.g., processes, files, and sockets), and edges encode interactions among them. 
Provenance graphs thus provide a rich temporal record of system behavior, making them well-suited for detecting sophisticated attacks such as advanced persistent threats (APTs)~\cite{milajerdi2019holmes,unicorn:han:2020}.

This representation has motivated many provenance-based intrusion detection systems (PIDSs) that apply machine learning---especially graph neural networks (GNNs)---to learn attack patterns from provenance data~\cite{Han2020SIGLSS,Wang2021THREATRACEDA,Li2023NODLINKAO,Jia2023MAGICDA,Cheng2023KairosPI,Rehman2024FlashAC,Goyal2024RCAIDER,jian2025,bilot2025simpler}. 
Despite reporting strong detection results, many systems face practical barriers that limit real-world adoption and hinder reproducibility~\cite{bilot2025simpler,abrar2025rep,dong2023we}.

A primary challenge is \emph{inconsistent evaluation practices}.
Even on the same datasets, evaluations differ in preprocessing, graph construction, and feature extraction. 
Dataset splits also lack a shared protocol: prior work varies temporal boundaries, hosts, and the mix of attack scenarios.
Ground-truth labeling is similarly inconsistent. 
Some approaches label entire neighborhoods as malicious, others label descendants of known malicious events, and others label specific nodes or edges.
Those choices that can substantially inflate or depress reported metrics~\cite{jian2025}. 
A recent reproducibility study~\cite{abrar2025rep} corroborates these concerns, reporting that many published results could not be reproduced due to missing code, data, or critical implementation details.

A second challenge is the \emph{high cost of re-implementation}.
New PIDSs are often built from scratch, requiring repeated effort to parse and preprocess provenance data, implement graph construction and feature-extraction pipelines, develop training and inference code, write evaluation scripts, and re-integrate baselines. 
This duplication is wasteful and error-prone: small implementation or tuning differences can skew results, potentially leading to unfair comparisons due to implementation issues or insufficiently tuning~\cite{arp2022and,bilot2025simpler}.

Prior work has also raised methodological concerns beyond evaluation inconsistency and engineering overhead~\cite{bilot2025simpler}.
Neural network--based systems can exhibit substantial run-to-run variability under different random seeds~\cite{bhojanapalli2021reproducibility,summers2021nondeterminism}, yet most studies report single-run results. Many approaches rely on manually tuned thresholds or clustering procedures that implicitly assume attack presence, introducing a form of data snooping~\cite{arp2022and}. 
Detection granularity also varies widely---some systems flag nodes or edges, others label neighborhoods or entire graphs---where coarser granularity can inflate metrics while providing less actionable outputs~\cite{jian2025,dong2023we}. 
Finally, PIDS performance is often sensitive to hyperparameters, but evaluation baselines are not always tuned with the same rigor as the proposed method.

To address these issues, we present \system, a unified framework for developing and evaluating PIDSs. 
\system integrates eight state-of-the-art systems within a modular, component-based architecture that supports consistent evaluation, rapid prototyping, and rigorous experimentation. 
It provides standardized preprocessing pipelines and ground-truth labels, removing major sources of evaluation inconsistency. 
A YAML-based configuration interface lets researchers assemble new PIDSs by mixing and matching components from existing systems, reducing development time and enabling systematic design-space exploration. 
Built-in support for ablation studies, hyperparameter tuning, stability analysis, and visualization further addresses the methodological gaps identified in prior work.

\noindgras{Contributions.}

\begin{itemize}[leftmargin=*]
	\item \textbf{Unified architecture.} We consolidate eight state-of-the-art PIDSs \cite{Wang2021THREATRACEDA, Li2023NODLINKAO, Jia2023MAGICDA, Cheng2023KairosPI, Rehman2024FlashAC, Goyal2024RCAIDER, jian2025, bilot2025simpler} into a single codebase with a modular, component-based architecture for consistent evaluation and fair comparison.
	\item \textbf{Standardized preprocessing and ground truth.} We provide standardized pipelines for parsing the DARPA TC~\cite{darpa-e3, darpa-e5} and OpTC~\cite{darpa-optc} datasets, along with consistent ground-truth labels following Jiang et al.~\cite{jian2025}, reducing major sources of evaluation inconsistency.
	\item \textbf{Config-driven prototyping.} Our YAML-based configuration system enables researchers to assemble new PIDSs by mixing and matching existing components without writing code, reducing development time and supporting systematic design-space exploration.
	\item \textbf{Experimental utilities.} We include integrated support for ablation studies, hyperparameter tuning, instability measurement, and visualization, addressing key methodological gaps identified in prior work~\cite{bilot2025simpler, arp2022and}.
	\item \textbf{Open-source release.} We release \system as open-source software, together with preprocessed datasets and ground-truth labels, to serve as a common evaluation platform for the community and encourage contributions to the framework.
\end{itemize}

%% file: fig/archi.tex
\begin{figure*}[t]
\centering
\resizebox{\textwidth}{!}{%
\begin{tikzpicture}[
    block/.style={
        rectangle,
        draw=black!70,
        fill=#1,
        minimum height=0.5cm,
        align=center,
        font=\tiny\bfseries,
        rounded corners=2pt
    },
    block/.default=blue!15,
    smallblock/.style={
        rectangle,
        draw=black!40,
        fill=#1,
        minimum width=1.2cm,
        minimum height=0.32cm,
        align=center,
        font=\tiny,
        rounded corners=1.5pt
    },
    smallblock/.default=gray!10,
    data/.style={
        cylinder,
        draw=black!70,
        fill=gray!25,
        shape border rotate=90,
        aspect=0.35,
        minimum width=0.9cm,
        minimum height=0.55cm,
        font=\scriptsize
    },
    arrow/.style={->, >=Stealth, thick, black!70},
    grouparrow/.style={->, >=Stealth, semithick, black!70},
    configarrow/.style={->, >=Stealth, dashed, black!40},
    brace/.style={decorate, decoration={brace, amplitude=5pt}},
    groupbox/.style={
        rectangle,
        draw=black!50,
        line width=0.6pt,
        rounded corners=6pt,
        inner sep=0pt
    },
    columnlabel/.style={
        font=\tiny\bfseries,
        text=black!60
    }
]

\definecolor{graphBlueFill}{RGB}{220,235,245}
\definecolor{transYellowFill}{RGB}{255,250,220}
\definecolor{featGreenFill}{RGB}{225,245,225}
\definecolor{batchPinkFill}{RGB}{255,235,240}
\definecolor{trainGrayFill}{RGB}{242,242,245}
\definecolor{detectPurpleFill}{RGB}{245,235,250}
\definecolor{triageLimeFill}{RGB}{235,250,225}

\definecolor{graphBlueBlock}{RGB}{180,210,230}
\definecolor{transYellowBlock}{RGB}{245,235,180}
\definecolor{featGreenBlock}{RGB}{195,225,195}
\definecolor{batchPinkBlock}{RGB}{245,205,215}
\definecolor{trainGrayBlock}{RGB}{215,215,220}
\definecolor{detectPurpleBlock}{RGB}{220,200,235}
\definecolor{triageLimeBlock}{RGB}{205,230,190}

\definecolor{graphBlueSmall}{RGB}{200,222,238}
\definecolor{transYellowSmall}{RGB}{248,242,200}
\definecolor{featGreenSmall}{RGB}{210,235,210}
\definecolor{batchPinkSmall}{RGB}{248,220,228}
\definecolor{trainTealSmall}{RGB}{205,230,230}
\definecolor{trainOrangeSmall}{RGB}{248,235,215}
\definecolor{trainRedSmall}{RGB}{245,222,222}
\definecolor{detectPurpleSmall}{RGB}{232,215,242}
\definecolor{detectCyanSmall}{RGB}{215,235,242}
\definecolor{triageLimeSmall}{RGB}{220,238,208}
\definecolor{noneGray}{RGB}{230,230,230}

\definecolor{metricsBlue}{RGB}{200,220,240}
\definecolor{vizBlue}{RGB}{100,150,200}

\def\vs{0.38}
\def\hs{2.0}
\def\tcol{1.45}
\def\tgap{0.44}
\def\startx{0}
\def\blockY{0.5}
\def\labelY{-0.15}
\def\optionStartY{-0.55}
\def\bottomPad{0.32}
\def\triageExtra{0.15}

\def\figureLeft{{\startx-0.78}}
\def\figureRight{{\startx+4*\hs+3*\tcol+3.52+2*\tgap+\triageExtra}}
\pgfmathsetmacro{\figureCenter}{(\startx-0.78 + \startx+4*\hs+3*\tcol+3.52+2*\tgap+\triageExtra)/2}

\pgfmathsetmacro{\trainCenter}{\startx+4*\hs+1.5*\tcol-0.24}

\pgfmathsetmacro{\triageGroupCenter}{(\startx+4*\hs+3*\tcol+2.22+2*\tgap + \startx+4*\hs+3*\tcol+3.52+2*\tgap+\triageExtra)/2}

\pgfmathsetmacro{\detectCenter}{\startx+4*\hs+3*\tcol+0.87+\tgap}

\node[data] (darpa) at ({\startx}, 2.0) {Datasets};

\begin{scope}[shift={(\figureCenter, 2.0)}]
    \fill[gray!15] (-0.5, -0.38) -- (-0.5, 0.38) -- (0.32, 0.38) -- (0.5, 0.2) -- (0.5, -0.38) -- cycle;
    \draw[black!70] (-0.5, -0.38) -- (-0.5, 0.38) -- (0.32, 0.38) -- (0.5, 0.2) -- (0.5, -0.38) -- cycle;
    \fill[gray!35] (0.32, 0.38) -- (0.32, 0.2) -- (0.5, 0.2) -- cycle;
    \draw[black!70] (0.32, 0.38) -- (0.32, 0.2) -- (0.5, 0.2);
    \node[font=\scriptsize, align=center] at (0, 0) {YAML\\Config};
\end{scope}

\begin{scope}[shift={({\triageGroupCenter-0.6}, 2.0)}]
    \fill[metricsBlue!50] (-0.42, -0.26) rectangle (0.42, 0.26);
    \draw[black!70] (-0.42, -0.26) rectangle (0.42, 0.26);
    \draw[metricsBlue!80!black] (-0.42, 0.087) -- (0.42, 0.087);
    \draw[metricsBlue!80!black] (-0.42, -0.087) -- (0.42, -0.087);
    \draw[metricsBlue!80!black] (0, -0.26) -- (0, 0.26);
    \node[font=\scriptsize, above] at (0, 0.32) {Metrics};
\end{scope}

\begin{scope}[shift={({\triageGroupCenter+0.6}, 2.0)}]
    \fill[metricsBlue!30] (-0.42, -0.26) rectangle (0.42, 0.26);
    \draw[black!70] (-0.42, -0.26) rectangle (0.42, 0.26);
    \fill[vizBlue!50] (-0.28, -0.16) rectangle (-0.12, 0.0);
    \fill[vizBlue!70] (-0.02, -0.16) rectangle (0.14, 0.1);
    \fill[vizBlue!90] (0.20, -0.16) rectangle (0.36, -0.04);
    \draw[vizBlue!70!black, thick] (-0.32, -0.16) -- (-0.32, 0.18);
    \draw[vizBlue!70!black, thick] (-0.32, -0.16) -- (0.38, -0.16);
    \node[font=\scriptsize, above] at (0, 0.32) {Visualization};
\end{scope}

\begin{scope}[on background layer]
    \fill[graphBlueFill, rounded corners=6pt] 
        ({\startx-0.78}, 0.95) rectangle ({\startx+0.78}, {\optionStartY-5*\vs-\bottomPad});
    \draw[groupbox] 
        ({\startx-0.78}, 0.95) rectangle ({\startx+0.78}, {\optionStartY-5*\vs-\bottomPad});
    
    \fill[transYellowFill, rounded corners=6pt] 
        ({\startx+\hs-0.78}, 0.95) rectangle ({\startx+\hs+0.78}, {\optionStartY-3*\vs-\bottomPad});
    \draw[groupbox] 
        ({\startx+\hs-0.78}, 0.95) rectangle ({\startx+\hs+0.78}, {\optionStartY-3*\vs-\bottomPad});
    
    \fill[featGreenFill, rounded corners=6pt] 
        ({\startx+2*\hs-0.78}, 0.95) rectangle ({\startx+2*\hs+0.78}, {\optionStartY-9*\vs-\bottomPad});
    \draw[groupbox] 
        ({\startx+2*\hs-0.78}, 0.95) rectangle ({\startx+2*\hs+0.78}, {\optionStartY-9*\vs-\bottomPad});
    
    \fill[batchPinkFill, rounded corners=6pt] 
        ({\startx+3*\hs-0.78}, 0.95) rectangle ({\startx+3*\hs+0.78}, {\optionStartY-2*\vs-\bottomPad});
    \draw[groupbox] 
        ({\startx+3*\hs-0.78}, 0.95) rectangle ({\startx+3*\hs+0.78}, {\optionStartY-2*\vs-\bottomPad});
    
    \fill[trainGrayFill, rounded corners=6pt] 
        ({\startx+4*\hs-0.78}, 0.95) rectangle ({\startx+4*\hs+3*\tcol-0.48}, {\optionStartY-10*\vs-\bottomPad});
    \draw[groupbox] 
        ({\startx+4*\hs-0.78}, 0.95) rectangle ({\startx+4*\hs+3*\tcol-0.48}, {\optionStartY-10*\vs-\bottomPad});
    
    \fill[detectPurpleFill, rounded corners=6pt] 
        ({\startx+4*\hs+3*\tcol-0.48+\tgap}, 0.95) rectangle ({\startx+4*\hs+3*\tcol+2.22+\tgap}, {\optionStartY-5*\vs-\bottomPad});
    \draw[groupbox] 
        ({\startx+4*\hs+3*\tcol-0.48+\tgap}, 0.95) rectangle ({\startx+4*\hs+3*\tcol+2.22+\tgap}, {\optionStartY-5*\vs-\bottomPad});
    
    \fill[triageLimeFill, rounded corners=6pt] 
        ({\startx+4*\hs+3*\tcol+2.22+2*\tgap}, 0.95) rectangle ({\startx+4*\hs+3*\tcol+3.52+2*\tgap+\triageExtra}, {\optionStartY-\vs-\bottomPad});
    \draw[groupbox] 
        ({\startx+4*\hs+3*\tcol+2.22+2*\tgap}, 0.95) rectangle ({\startx+4*\hs+3*\tcol+3.52+2*\tgap+\triageExtra}, {\optionStartY-\vs-\bottomPad});
\end{scope}

\draw[grouparrow] (darpa.south) -- (\startx, 0.95);

\draw[grouparrow] ({\startx+0.78}, \blockY) -- ({\startx+\hs-0.78}, \blockY);
\draw[grouparrow] ({\startx+\hs+0.78}, \blockY) -- ({\startx+2*\hs-0.78}, \blockY);
\draw[grouparrow] ({\startx+2*\hs+0.78}, \blockY) -- ({\startx+3*\hs-0.78}, \blockY);
\draw[grouparrow] ({\startx+3*\hs+0.78}, \blockY) -- ({\startx+4*\hs-0.78}, \blockY);
\draw[grouparrow] ({\startx+4*\hs+3*\tcol-0.48}, \blockY) -- ({\startx+4*\hs+3*\tcol-0.48+\tgap}, \blockY);
\draw[grouparrow] ({\startx+4*\hs+3*\tcol+2.22+\tgap}, \blockY) -- ({\startx+4*\hs+3*\tcol+2.22+2*\tgap}, \blockY);

\coordinate (triageTop) at (\triageGroupCenter, 0.95);
\coordinate (metricsBot) at ({\triageGroupCenter-0.6}, 1.48);
\coordinate (vizBot) at ({\triageGroupCenter+0.6}, 1.48);
\draw[grouparrow] (triageTop) -- ++(0, 0.25) -| (metricsBot);
\draw[grouparrow] (triageTop) -- ++(0, 0.25) -| (vizBot);

\coordinate (yamlbase) at (\figureCenter, 1.62);
\draw[configarrow] (yamlbase) -- (\startx, 0.95);
\draw[configarrow] (yamlbase) -- ({\startx+\hs}, 0.95);
\draw[configarrow] (yamlbase) -- ({\startx+2*\hs}, 0.95);
\draw[configarrow] (yamlbase) -- ({\startx+3*\hs}, 0.95);
\draw[configarrow] (yamlbase) -- (\trainCenter, 0.95);
\draw[configarrow] (yamlbase) -- (\detectCenter, 0.95);
\draw[configarrow] (yamlbase) -- (\triageGroupCenter, 0.95);

\node[block=graphBlueBlock, minimum width=1.3cm] at (\startx, \blockY) {Construction};
\node[columnlabel] at (\startx, \labelY) {Features};

\node[block=transYellowBlock, minimum width=1.3cm] at ({\startx+\hs}, \blockY) {Transformation};
\node[columnlabel] at ({\startx+\hs}, \labelY) {Graph Transform.};

\node[block=featGreenBlock, minimum width=1.3cm] at ({\startx+2*\hs}, \blockY) {Featurization};
\node[columnlabel] at ({\startx+2*\hs}, \labelY) {Text Embedding};

\node[block=batchPinkBlock, minimum width=1.3cm] at ({\startx+3*\hs}, \blockY) {Batching};
\node[columnlabel] at ({\startx+3*\hs}, \labelY) {Graph Batching};

\pgfmathsetmacro{\trainGroupCenter}{(\startx+4*\hs-0.78 + \startx+4*\hs+3*\tcol-0.48)/2}
\node[block=trainGrayBlock, minimum width=3.6cm] at (\trainGroupCenter, \blockY) {Training};
\node[columnlabel] at ({\startx+4*\hs}, \labelY) {Encoder};
\node[columnlabel] at ({\startx+4*\hs+\tcol}, \labelY) {Decoder};
\node[columnlabel] at ({\startx+4*\hs+2*\tcol}, \labelY) {Objective};

\pgfmathsetmacro{\detectGroupCenter}{(\startx+4*\hs+3*\tcol-0.48+\tgap + \startx+4*\hs+3*\tcol+2.22+\tgap)/2}
\node[block=detectPurpleBlock, minimum width=2.4cm] at (\detectGroupCenter, \blockY) {Evaluation};
\node[columnlabel] at ({\startx+4*\hs+3*\tcol+0.22+\tgap}, \labelY) {Threshold};
\node[columnlabel] at ({\startx+4*\hs+3*\tcol+1.52+\tgap}, \labelY) {Detection};

\node[block=triageLimeBlock, minimum width=1.15cm] at (\triageGroupCenter, \blockY) {Triage};
\node[columnlabel] at (\triageGroupCenter, \labelY) {Tracing};

\node[smallblock=graphBlueSmall] at (\startx, \optionStartY) {Type};
\node[smallblock=graphBlueSmall] at (\startx, {\optionStartY-\vs}) {File Path};
\node[smallblock=graphBlueSmall] at (\startx, {\optionStartY-2*\vs}) {Process Path};
\node[smallblock=graphBlueSmall] at (\startx, {\optionStartY-3*\vs}) {Process Cmd};
\node[smallblock=graphBlueSmall] at (\startx, {\optionStartY-4*\vs}) {Socket IP};
\node[smallblock=graphBlueSmall] at (\startx, {\optionStartY-5*\vs}) {Socket Port};

\node[smallblock=transYellowSmall] at ({\startx+\hs}, \optionStartY) {R-CAID Pseudo};
\node[smallblock=transYellowSmall] at ({\startx+\hs}, {\optionStartY-\vs}) {Undirected};
\node[smallblock=transYellowSmall] at ({\startx+\hs}, {\optionStartY-2*\vs}) {DAG};
\node[smallblock=noneGray] at ({\startx+\hs}, {\optionStartY-3*\vs}) {None};

\node[smallblock=featGreenSmall] at ({\startx+2*\hs}, \optionStartY) {Word2Vec};
\node[smallblock=featGreenSmall] at ({\startx+2*\hs}, {\optionStartY-\vs}) {Doc2Vec};
\node[smallblock=featGreenSmall] at ({\startx+2*\hs}, {\optionStartY-2*\vs}) {FastText};
\node[smallblock=featGreenSmall] at ({\startx+2*\hs}, {\optionStartY-3*\vs}) {ALaCarte};
\node[smallblock=featGreenSmall] at ({\startx+2*\hs}, {\optionStartY-4*\vs}) {Temporal RW};
\node[smallblock=featGreenSmall] at ({\startx+2*\hs}, {\optionStartY-5*\vs}) {Flash};
\node[smallblock=featGreenSmall] at ({\startx+2*\hs}, {\optionStartY-6*\vs}) {HFH};
\node[smallblock=featGreenSmall] at ({\startx+2*\hs}, {\optionStartY-7*\vs}) {MAGIC};
\node[smallblock=featGreenSmall] at ({\startx+2*\hs}, {\optionStartY-8*\vs}) {Only Type};
\node[smallblock=featGreenSmall] at ({\startx+2*\hs}, {\optionStartY-9*\vs}) {Only Ones};

\node[smallblock=batchPinkSmall] at ({\startx+3*\hs}, \optionStartY) {Global};
\node[smallblock=batchPinkSmall] at ({\startx+3*\hs}, {\optionStartY-\vs}) {Intra-graph};
\node[smallblock=batchPinkSmall] at ({\startx+3*\hs}, {\optionStartY-2*\vs}) {Inter-graph};

\node[smallblock=trainTealSmall] at ({\startx+4*\hs}, \optionStartY) {TGN};
\node[smallblock=trainTealSmall] at ({\startx+4*\hs}, {\optionStartY-\vs}) {Graph Attention};
\node[smallblock=trainTealSmall] at ({\startx+4*\hs}, {\optionStartY-2*\vs}) {SAGE};
\node[smallblock=trainTealSmall] at ({\startx+4*\hs}, {\optionStartY-3*\vs}) {GAT};
\node[smallblock=trainTealSmall] at ({\startx+4*\hs}, {\optionStartY-4*\vs}) {GIN};
\node[smallblock=trainTealSmall] at ({\startx+4*\hs}, {\optionStartY-5*\vs}) {Sum Aggregation};
\node[smallblock=trainTealSmall] at ({\startx+4*\hs}, {\optionStartY-6*\vs}) {R-CAID GAT};
\node[smallblock=trainTealSmall] at ({\startx+4*\hs}, {\optionStartY-7*\vs}) {MAGIC GAT};
\node[smallblock=trainTealSmall] at ({\startx+4*\hs}, {\optionStartY-8*\vs}) {GLSTM};
\node[smallblock=trainTealSmall] at ({\startx+4*\hs}, {\optionStartY-9*\vs}) {Custom MLP};
\node[smallblock=noneGray] at ({\startx+4*\hs}, {\optionStartY-10*\vs}) {None};

\node[smallblock=trainOrangeSmall] at ({\startx+4*\hs+\tcol}, \optionStartY) {Edge MLP};
\node[smallblock=trainOrangeSmall] at ({\startx+4*\hs+\tcol}, {\optionStartY-\vs}) {Node MLP};
\node[smallblock=trainOrangeSmall] at ({\startx+4*\hs+\tcol}, {\optionStartY-2*\vs}) {MAGIC GAT};
\node[smallblock=trainOrangeSmall] at ({\startx+4*\hs+\tcol}, {\optionStartY-3*\vs}) {NodLink};
\node[smallblock=trainOrangeSmall] at ({\startx+4*\hs+\tcol}, {\optionStartY-4*\vs}) {Inner Product};
\node[smallblock=noneGray] at ({\startx+4*\hs+\tcol}, {\optionStartY-5*\vs}) {None};

\node[smallblock=trainRedSmall] at ({\startx+4*\hs+2*\tcol}, \optionStartY) {Edge Type};
\node[smallblock=trainRedSmall] at ({\startx+4*\hs+2*\tcol}, {\optionStartY-\vs}) {Node Type};
\node[smallblock=trainRedSmall] at ({\startx+4*\hs+2*\tcol}, {\optionStartY-2*\vs}) {Masked Struct};
\node[smallblock=trainRedSmall] at ({\startx+4*\hs+2*\tcol}, {\optionStartY-3*\vs}) {Few-Shot Edge};
\node[smallblock=trainRedSmall] at ({\startx+4*\hs+2*\tcol}, {\optionStartY-4*\vs}) {Contrastive};
\node[smallblock=trainRedSmall] at ({\startx+4*\hs+2*\tcol}, {\optionStartY-5*\vs}) {Node Feat Recon};
\node[smallblock=trainRedSmall] at ({\startx+4*\hs+2*\tcol}, {\optionStartY-6*\vs}) {Node Emb Recon};
\node[smallblock=trainRedSmall] at ({\startx+4*\hs+2*\tcol}, {\optionStartY-7*\vs}) {Edge Emb Recon};
\node[smallblock=trainRedSmall] at ({\startx+4*\hs+2*\tcol}, {\optionStartY-8*\vs}) {Masked Feat Recon};

\node[smallblock=detectPurpleSmall] at ({\startx+4*\hs+3*\tcol+0.22+\tgap}, \optionStartY) {Max Val Loss};
\node[smallblock=detectPurpleSmall] at ({\startx+4*\hs+3*\tcol+0.22+\tgap}, {\optionStartY-\vs}) {Mean Val Loss};
\node[smallblock=detectPurpleSmall] at ({\startx+4*\hs+3*\tcol+0.22+\tgap}, {\optionStartY-2*\vs}) {ThreaTrace};
\node[smallblock=detectPurpleSmall] at ({\startx+4*\hs+3*\tcol+0.22+\tgap}, {\optionStartY-3*\vs}) {MAGIC};
\node[smallblock=detectPurpleSmall] at ({\startx+4*\hs+3*\tcol+0.22+\tgap}, {\optionStartY-4*\vs}) {Flash};
\node[smallblock=detectPurpleSmall] at ({\startx+4*\hs+3*\tcol+0.22+\tgap}, {\optionStartY-5*\vs}) {NodLink};

\node[smallblock=detectCyanSmall] at ({\startx+4*\hs+3*\tcol+1.52+\tgap}, \optionStartY) {Node Level};
\node[smallblock=detectCyanSmall] at ({\startx+4*\hs+3*\tcol+1.52+\tgap}, {\optionStartY-\vs}) {Edge Level};
\node[smallblock=detectCyanSmall] at ({\startx+4*\hs+3*\tcol+1.52+\tgap}, {\optionStartY-2*\vs}) {Time Window};

\node[smallblock=triageLimeSmall] at (\triageGroupCenter, \optionStartY) {DepImpact};
\node[smallblock=noneGray] at (\triageGroupCenter, {\optionStartY-\vs}) {None};

\begin{scope}[shift={({\startx+4*\hs+3*\tcol+1.5+\tgap}, -4.6)}]
    \draw[grouparrow] (0, 0.25) -- (0.7, 0.25) node[anchor=west, font=\tiny] {Data flow};
    \draw[configarrow] (0, -0.05) -- (0.7, -0.05) node[anchor=west, font=\tiny] {Configuration};
\end{scope}

\draw[brace, black!60] ({\startx+4*\hs+3*\tcol+3.60+2*\tgap+\triageExtra}, -5.05) -- ({\startx-0.85}, -5.05) 
    node[midway, below=5pt, font=\scriptsize, text=black!70] {Modular pipeline with on-disk caching and automatic restarting};

\end{tikzpicture}%
}
\caption{\textsc{PIDSMaker} architecture with seven pipeline stages connected by data flow. Each stage contains configurable components.}
\label{fig:architecture}
\end{figure*}
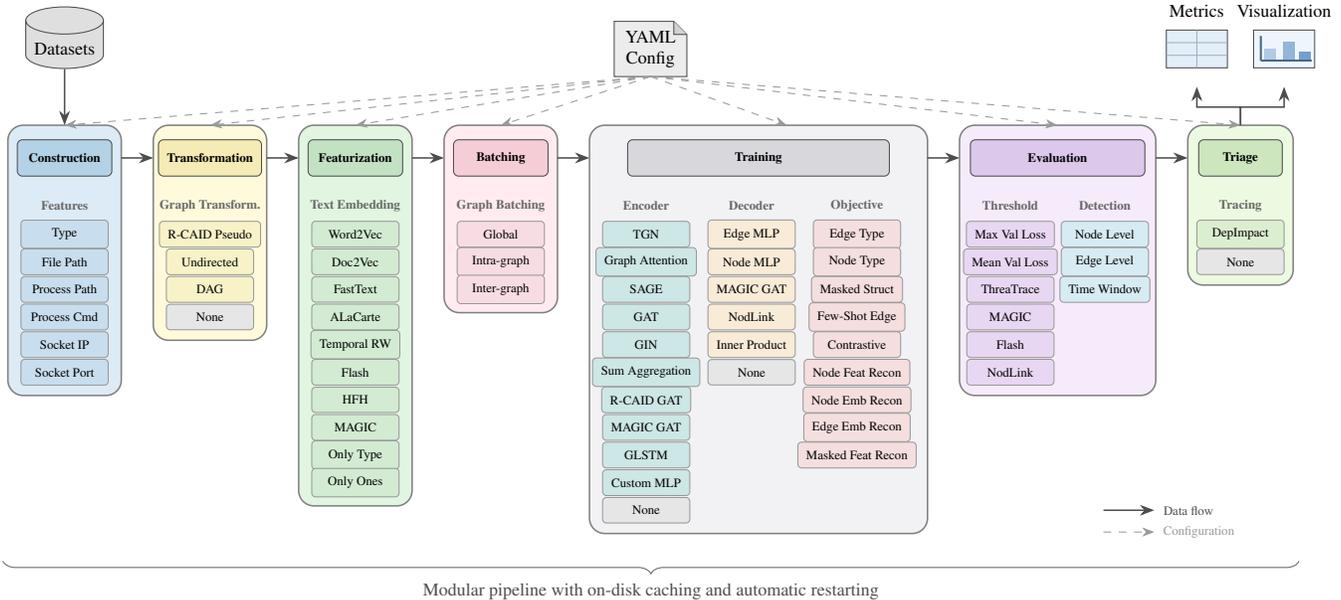

%% file: sections/background.tex
We review provenance graphs and the common architectural patterns of PIDSs that inform the design of \system.

\subsection{Provenance Graphs}

A provenance graph $G = (V, E)$ is a directed graph that captures causal relationships among system entities during execution. Nodes $v \in V$ represent entities such as processes, files, and network sockets, while edges $e \in E$ represent interactions between entities, such as a process reading from a file or writing to a socket.

Nodes and edges carry attributes that provide semantic context. For example, a process node may include the executable path and command-line arguments. Edges are typically annotated with the system-call type (e.g., \texttt{read}, \texttt{write}, \texttt{execute}) and a timestamp indicating when the interaction occurred.

Provenance graphs are derived from system audit logs that record low-level events. 
Capture mechanisms include kernel-level auditing~\cite{pohly2012hi,Bates2015TrustworthyWP, Pasquier2017PracticalWP}, dynamic instrumentation~\cite{Stamatogiannakis2014LookingIT}, and hardware-assisted tracing~\cite{Thalheim2016INSPECTORDP, Zeng2022PalanTrOA}. 
In production environments, the resulting graphs can grow rapidly, accumulating millions of nodes and edges over time.

\subsection{Common PIDS Architecture}

Despite their diversity, modern machine learning-based PIDSs share a common high-level pipeline. Raw provenance is first \emph{parsed and preprocessed} to extract attributes and construct graph structure. 
The graph may then undergo \emph{transformations} such as edge reduction or conversion to directed acyclic graphs. \emph{Feature extraction} converts textual and categorical attributes into numerical representations, and the data is \emph{batched} into temporal subgraphs for efficient processing.

During \emph{training}, benign activity is fed to an encoder (typically a GNN) to produce node or edge embeddings, which a decoder maps to predictions. 
Training commonly uses self-supervised objectives such as edge-type prediction, node-type prediction, or reconstruction. 
At \emph{inference}, anomaly scores are computed from prediction errors and thresholded to produce detection decisions. 
Evaluation compares flagged entities against ground-truth labels provided by existing datasets. Finally, an optional \emph{triage} stage prioritizes anomalies to reduce analyst workload.

Although this architecture suggests that components should be interchangeable, systems implement these stages independently, often with subtle but consequential differences. 
Such variations---together with inconsistent preprocessing, labeling, and evaluation protocols---complicate fair comparison and motivate the unified framework presented in the next section.

%% file: sections/design.tex
We describe the design and architecture of \system (\autoref{fig:architecture}) by outlining the core design principles, presenting the modular pipeline, and summarizing key implementation details, including caching, datasets, and supported systems.
The framework is implemented in Python using PyTorch and PyTorch Geometric~\cite{pyg}.

\subsection{Design Principles}

\noindent We follow four design principles:

\noindgras{Modularity.} Each PIDS stage is an independent, interchangeable component. 
This enables mixing components across systems and supports ablation studies. 
As shown in Figure~\ref{fig:modularity}, components can be freely combined to create new variants.

\noindgras{Configurability.} Systems are specified in YAML rather than code. 
This improves reproducibility and lowers the barrier to entry: a complete PIDS can be declared without writing code.

\noindgras{Efficiency.} \system avoids redundant computation via on-disk caching keyed by configuration.
Each stage caches its outputs in a uniquely hashed directory, enabling rapid iteration when only downstream settings change.

\noindgras{Extensibility.} New components can be added by implementing simple interfaces, allowing the \system to evolve and easily incorporate new systems.

\subsection{Pipeline Stages}

Based on our analysis of existing systems, we model the detection pipeline as seven stages, illustrated in \autoref{fig:architecture}). 
The figure also summarizes the implementations currently available in \system across these stages, including components reimplemented from public PIDS repositories and additional components that enable more advanced variants.

\begin{enumerate}[leftmargin=*]
	\item \textbf{Construction.} Parse raw provenance to construct a graph and extract attributes (e.g., entity types, file paths, process command lines, and network addresses).
	
	\item \textbf{Transformation.} Apply graph transformations to improve learning, such as converting to undirected graphs, removing redundant edges, converting to directed acyclic graphs (DAGs), or constructing pseudo-graphs that connect nodes to root causes.
	
	\item \textbf{Featurization.} Convert attributes into numerical representations using text embeddings (e.g., Word2Vec~\cite{mikolov2013efficient}, Doc2Vec~\cite{le2014distributed}, FastText~\cite{bojanowski2017enriching}) or domain-specific encodings such as Hierarchical Feature Hashing (HFH)~\cite{zhang2020dynamic}.
	
	\item \textbf{Batching.} Partition provenance into temporal subgraphs that fit in memory, typically using fixed-duration windows or a fixed number of events.
	
	\item \textbf{Training.} Train an encoder (typically a GNN) to produce node/edge embeddings and a decoder to map embeddings to predictions, using a self-supervised objective over benign activity.
	
	\item \textbf{Evaluation.} At inference, compute anomaly scores from prediction errors or reconstruction losses and threshold them (e.g., fixed thresholds, validation-based thresholds, or clustering) to produce detections. Metrics may be reported at node-, edge-, or window-level granularity, which can substantially affect results.
	
	\item \textbf{Triage.} Optionally prioritize detections and identify root causes using techniques such as DepImpact~\cite{fang2022back}.
\end{enumerate}

\subsection{On-Disk Caching}

Each stage is parameterized by YAML-defined arguments (\eg hyperparameters, number of workers). 
To avoid redundant computation, \system computes a unique hash for each stage from its arguments and its predecessor's hash, and stores outputs under a directory named by this hash. 
Before executing a stage, the pipeline checks for the corresponding directory; if it exists, the stage is skipped and cached results are loaded.

This design supports efficient iteration.
For example, changing only the learning rate reuses cached outputs from upstream stages (construction, transformation, featurization, batching) and re-runs only the affected stages.

\subsection{Datasets and Ground Truth}

\system includes preprocessed versions of standard DARPA datasets used in PIDS research:
\begin{itemize}[leftmargin=*]
	\item \textbf{DARPA TC E3}~\cite{darpa-e3}: five hosts (CADETS, ClearScope, THEIA, FiveDirections, Trace) with multiple attacks each.
	\item \textbf{DARPA TC E5}~\cite{darpa-e5}: five hosts with varying numbers of attacks.
	\item \textbf{DARPA OpTC}~\cite{darpa-optc}: three Windows hosts (H051, H201, H501) with one attack each.
\end{itemize}

In total, \system integrates 13 datasets: \texttt{CADETS\_E3}, \texttt{THEIA\_E3}, \texttt{CLEARSCOPE\_E3}, \texttt{FIVEDIRECTIONS\_E3}, \texttt{TRACE\_E3}, \texttt{CADETS\_E5}, \texttt{THEIA\_E5}, \texttt{CLEARSCOPE\_E5}, \texttt{FIVEDIRECTIONS\_E5}, \texttt{TRACE\_E5}, \texttt{optc\_h201}, \texttt{optc\_h501}, and \texttt{optc\_h051}.

We provide consistent ground-truth labels following Jiang et al.~\cite{jian2025}, focusing on node-level detection rather than neighborhood- or graph-level labeling. These labels are available for download and enable consistent evaluation across supported systems.

\subsection{Supported Systems}

\system currently implements eight state-of-the-art PIDSs spanning diverse design choices: \textsc{ThreaTrace}~\cite{Wang2021THREATRACEDA}, \textsc{NodLink}~\cite{Li2023NODLINKAO}, \textsc{Magic}~\cite{Jia2023MAGICDA}, \textsc{Kairos}~\cite{Cheng2023KairosPI}, \textsc{Flash}~\cite{Rehman2024FlashAC}, \textsc{R-Caid}~\cite{Goyal2024RCAIDER}, \textsc{Orthrus}~\cite{jian2025}, and \textsc{Velox}~\cite{bilot2025simpler}. Each system is specified by a YAML configuration that selects components and parameters. For systems without public source code (e.g., \textsc{R-Caid}), we provide re-implementations based on the papers.

\input{fig/modularity}

\subsection{Configuration Files}

Each system is defined by a YAML configuration file that specifies all pipeline components and their parameters. The following is an excerpt from the \textsc{Orthrus} configuration:

\begin{yamlcode}[title=orthrus.yml]
construction:
  node_features:
    subject: type, path, cmd_line
    file: type, path
    netflow: type, remote_ip, remote_port
    [...]
transformation:
  used_methods: none
featurization:
  used_method: word2vec
  emb_dim: 128
  epochs: 50
  word2vec:
    alpha: 0.025
    [...]
batching:
  intra_graph_batching:
    used_methods: edges, tgn_last_neighbor
    [...]
training:
  lr: 0.00001
  node_hid_dim: 128
  encoder:
    used_methods: tgn, graph_attention
    graph_attention:
      num_heads: 8
      [...]
  decoder:
    [...]
evaluation:
  used_method: node_evaluation
  node_evaluation:
    threshold_method: max_val_loss
    [...]
triage:
  used_method: depimpact
  use_kmeans: True
  [...]
\end{yamlcode}

CLI arguments can override any YAML parameter using dot notation (e.g., \texttt{--training.lr=0.0001}) and take precedence over YAML values. The \texttt{--force\_restart} flag re-executes stages from a specified point even when arguments are unchanged, while \texttt{--restart\_from\_scratch} writes outputs to a new directory to isolate runs.

\subsection{Minimum System Requirements}
\system is designed to train deep neural networks on large volumes of system data and is therefore primarily intended to run on servers equipped with substantial memory capacity and CPU resources.
We recommend a minimum of 100 GB of system memory for processing the smallest datasets, and at least 500 GB for the largest datasets in order to avoid resource constraints. To improve throughput, \system keeps most data resident in memory, thereby reducing time-consuming disk I/O operations, which in turn necessitates significant memory availability.
\system supports execution on both CPUs and GPUs. However, we strongly recommend the use of a GPU with at least 20 GB of memory for small to medium-scale datasets, and at least 40 GB for larger datasets.

%% file: fig/modularity.tex
\begin{figure*}[t]
\centering
\begin{tikzpicture}[
    component/.style={
        rectangle,
        draw=black!70,
        minimum width=1.7cm,
        minimum height=0.45cm,
        align=center,
        font=\tiny,
        rounded corners=1.5pt,
        inner sep=1pt
    },
    selectedcomp/.style={
        rectangle,
        draw=black,
        line width=0.8pt,
        minimum width=1.7cm,
        minimum height=0.45cm,
        align=center,
        font=\tiny,
        rounded corners=1.5pt,
        inner sep=1pt
    },
    emptycomp/.style={
        rectangle,
        draw=gray!40,
        minimum width=1.7cm,
        minimum height=0.45cm,
        align=center,
        font=\tiny,
        rounded corners=1.5pt,
        inner sep=1pt,
        fill=gray!5,
        text=gray!60
    },
    emptyselected/.style={
        rectangle,
        draw=black,
        line width=0.8pt,
        minimum width=1.7cm,
        minimum height=0.45cm,
        align=center,
        font=\tiny,
        rounded corners=1.5pt,
        inner sep=1pt,
        fill=gray!10,
        text=gray!60
    },
    sysname/.style={
        font=\scriptsize\scshape,
        anchor=east
    },
    colheader/.style={
        rectangle,
        draw=black,
        minimum width=1.7cm,
        minimum height=0.7cm,
        align=center,
        font=\scriptsize,
        rounded corners=2pt
    },
    arrow/.style={
        ->,
        >=stealth,
        gray!50
    },
    mixarrow/.style={
        -Triangle,
        line width=0.8pt,
        black!80
    }
]

\def\vs{0.52}

\def\colA{2.0}    
\def\colB{4.2}    
\def\colC{6.4}    
\def\colD{8.6}    
\def\colE{10.8}   
\def\colF{13.0}   
\def\colG{15.2}   

\def\rowA{4.0}    
\def\rowB{3.48}   
\def\rowC{2.96}   
\def\rowD{2.44}   
\def\rowE{1.92}   
\def\rowF{1.40}   
\def\rowG{0.88}   
\def\rowH{0.36}   

\node[colheader, fill=blue!25] at (\colA, 4.75) {Features};
\node[colheader, fill=yellow!35] at (\colB, 4.75) {Graph\\Transformation};
\node[colheader, fill=green!25] at (\colC, 4.75) {Text\\Embedding};
\node[colheader, fill=teal!30] at (\colD, 4.75) {Encoder};
\node[colheader, fill=orange!25] at (\colE, 4.75) {Decoder};
\node[colheader, fill=red!25] at (\colF, 4.75) {Objective};
\node[colheader, fill=purple!25] at (\colG, 4.75) {Threshold};

\node[sysname] at (1.05, \rowA) {ThreaTrace};
\node[sysname] at (1.05, \rowB) {NodLink};
\node[sysname] at (1.05, \rowC) {Magic};
\node[sysname] at (1.05, \rowD) {Kairos};
\node[sysname] at (1.05, \rowE) {Flash};
\node[sysname] at (1.05, \rowF) {R-Caid};
\node[sysname] at (1.05, \rowG) {Orthrus};
\node[sysname] at (1.05, \rowH) {Velox};

\node[component, fill=blue!10] (t1) at (\colA, \rowA) {type+edge dist};
\node[emptycomp] (t2) at (\colB, \rowA) {--};
\node[emptycomp] (t3) at (\colC, \rowA) {--};
\node[component, fill=teal!15] (t4) at (\colD, \rowA) {GraphSAGE};
\node[emptycomp] (t5) at (\colE, \rowA) {--};
\node[component, fill=red!10] (t6) at (\colF, \rowA) {Node Type};
\node[component, fill=purple!10] (t7) at (\colG, \rowA) {Fixed Thresh};

\node[component, fill=blue!10] (n1) at (\colA, \rowB) {cmd, path, IP};
\node[component, fill=yellow!15] (n2) at (\colB, \rowB) {Undirected};
\node[selectedcomp, fill=green!10] (n3) at (\colC, \rowB) {FastText};
\node[component, fill=teal!15] (n4) at (\colD, \rowB) {Weighted Sum};
\node[component, fill=orange!10] (n5) at (\colE, \rowB) {VAE+MLP};
\node[component, fill=red!10] (n6) at (\colF, \rowB) {Node Recon};
\node[component, fill=purple!10] (n7) at (\colG, \rowB) {Val Thresh};

\node[component, fill=blue!10] (m1) at (\colA, \rowC) {node+edge type};
\node[component, fill=yellow!15] (m2) at (\colB, \rowC) {No redundant};
\node[emptycomp] (m3) at (\colC, \rowC) {--};
\node[component, fill=teal!15] (m4) at (\colD, \rowC) {GAT};
\node[component, fill=orange!10] (m5) at (\colE, \rowC) {GAT+MLP};
\node[component, fill=red!10] (m6) at (\colF, \rowC) {Masked Recon};
\node[component, fill=purple!10] (m7) at (\colG, \rowC) {K-D Tree};

\node[component, fill=blue!10] (k1) at (\colA, \rowD) {path, IP+port};
\node[emptycomp] (k2) at (\colB, \rowD) {--};
\node[component, fill=green!10] (k3) at (\colC, \rowD) {HFH};
\node[component, fill=teal!15] (k4) at (\colD, \rowD) {TGN+Attn};
\node[component, fill=orange!10] (k5) at (\colE, \rowD) {MLP};
\node[component, fill=red!10] (k6) at (\colF, \rowD) {Edge Type};
\node[component, fill=purple!10] (k7) at (\colG, \rowD) {Fixed Thresh};

\node[component, fill=blue!10] (f1) at (\colA, \rowE) {cmd, path, IP};
\node[emptycomp] (f2) at (\colB, \rowE) {--};
\node[component, fill=green!10] (f3) at (\colC, \rowE) {W2V+PosEnc};
\node[selectedcomp, fill=teal!15] (f4) at (\colD, \rowE) {GraphSAGE};
\node[component, fill=orange!10] (f5) at (\colE, \rowE) {XGBoost};
\node[component, fill=red!10] (f6) at (\colF, \rowE) {Node Type};
\node[component, fill=purple!10] (f7) at (\colG, \rowE) {Fixed Thresh};

\node[component, fill=blue!10] (r1) at (\colA, \rowF) {path+name};
\node[component, fill=yellow!15] (r2) at (\colB, \rowF) {Pseudo-graph};
\node[component, fill=green!10] (r3) at (\colC, \rowF) {Doc2Vec};
\node[component, fill=teal!15] (r4) at (\colD, \rowF) {GAT};
\node[emptycomp] (r5) at (\colE, \rowF) {--};
\node[component, fill=red!10] (r6) at (\colF, \rowF) {Node Type};
\node[component, fill=purple!10] (r7) at (\colG, \rowF) {K-Means+MAD};

\node[selectedcomp, fill=blue!10] (o1) at (\colA, \rowG) {type,path,cmd,IP};
\node[emptyselected] (o2) at (\colB, \rowG) {--};
\node[component, fill=green!10] (o3) at (\colC, \rowG) {Word2Vec};
\node[component, fill=teal!15] (o4) at (\colD, \rowG) {TGN-Light};
\node[selectedcomp, fill=orange!10] (o5) at (\colE, \rowG) {MLP};
\node[component, fill=red!10] (o6) at (\colF, \rowG) {Edge Type};
\node[component, fill=purple!10] (o7) at (\colG, \rowG) {Val+K-Means};

\node[component, fill=blue!10] (v1) at (\colA, \rowH) {type,path,cmd,IP};
\node[emptycomp] (v2) at (\colB, \rowH) {--};
\node[component, fill=green!10] (v3) at (\colC, \rowH) {Word2Vec};
\node[component, fill=teal!15] (v4) at (\colD, \rowH) {Linear};
\node[component, fill=orange!10] (v5) at (\colE, \rowH) {MLP};
\node[selectedcomp, fill=red!10] (v6) at (\colF, \rowH) {Edge Type};
\node[selectedcomp, fill=purple!10] (v7) at (\colG, \rowH) {Val Thresh};

\foreach \col in {\colA, \colB, \colC, \colD, \colE, \colF} {
    \draw[gray!20] (\col+0.95, 4.35) -- (\col+0.95, 0.1);
}

\foreach \row in {\rowA, \rowB, \rowC, \rowD, \rowE, \rowF, \rowG, \rowH} {
    \draw[arrow] (\colA+0.9, \row) -- (\colB-0.9, \row);
    \draw[arrow] (\colB+0.9, \row) -- (\colC-0.9, \row);
    \draw[arrow] (\colC+0.9, \row) -- (\colD-0.9, \row);
    \draw[arrow] (\colD+0.9, \row) -- (\colE-0.9, \row);
    \draw[arrow] (\colE+0.9, \row) -- (\colF-0.9, \row);
    \draw[arrow] (\colF+0.9, \row) -- (\colG-0.9, \row);
}


\draw[mixarrow] (o1.east) -- (o2.west);

\draw[mixarrow, rounded corners=4pt] 
    (o2.east) -- (\colB+0.95, \rowG) -- (\colB+0.95, \rowB) -- (n3.west);

\draw[mixarrow, rounded corners=4pt] 
    (n3.east) -- (\colC+0.95, \rowB) -- (\colC+0.95, \rowE) -- (f4.west);

\draw[mixarrow, rounded corners=4pt] 
    (f4.east) -- (\colD+0.95, \rowE) -- (\colD+0.95, \rowG) -- (o5.west);

\draw[mixarrow, rounded corners=4pt] 
    (o5.east) -- (\colE+0.95, \rowG) -- (\colE+0.95, \rowH) -- (v6.west);

\draw[mixarrow] (v6.east) -- (v7.west);

\node[font=\scriptsize\bfseries, black!80, fill=white, inner sep=2pt] at (\colB+0.95, \rowD+0.15) {custom system};

\end{tikzpicture}
\caption{Component modularity in \textsc{PIDSMaker}. Rows represent the eight supported systems; columns represent a subset of components. The black arrow demonstrates creating a custom system variant by selecting components from different systems. This composition can be specified through YAML or CLI.}
\label{fig:modularity}
\end{figure*}
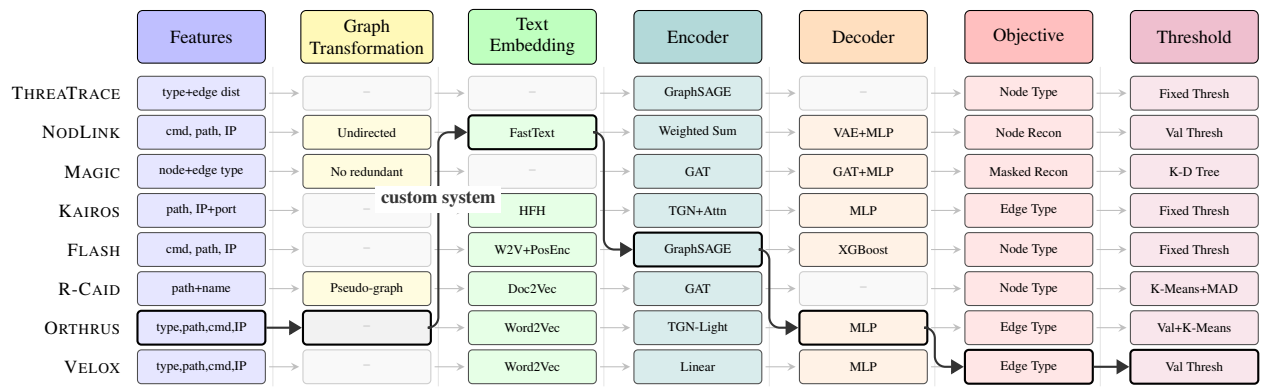

%% file: sections/usecase.tex
We illustrate the key capabilities of \system through concrete examples: running existing systems, defining new variants, performing ablations and hyperparameter tuning, measuring instability, and visualizing results.

\subsection{Running Existing Systems}

Running a system requires specifying a system's configuration and a dataset:

\begin{bashcode}[title=Terminal]
	./run.sh SYSTEM DATASET
\end{bashcode}

\texttt{SYSTEM} selects a YAML configuration in \texttt{config/}, and \texttt{DATASET} selects a supported dataset. 
The \texttt{run.sh} script executes the pipeline in the background by default, as the framework is designed to be monitored through Weights \& Biases (W\&B)~\cite{wandb}. Logs, metrics, and figures are streamed and uploaded to the platform in real time.
W\&B is a web-based platform for experiment tracking and model versioning, which interfaces with the codebase via API calls.
\system integrates W\&B by default; however, a local-only execution mode is also supported by removing the \texttt{--wandb} flag.
Examples of metrics and figures exported to the interface are provided in \autoref{sec:viz}.

By default, \system runs on GPU (CUDA:0) and falls back to CPU if no GPU is available; CPU execution can be forced with \texttt{--cpu}.

\subsection{Designing Model Variants}

\system enables rapid prototyping by composing pipeline components from existing systems (Figure~\ref{fig:modularity}). To define a new variant, users create a YAML configuration that inherits from a base system and overrides only the relevant fields. For example, the following configuration modifies Orthrus by changing the featurization method, encoder, and evaluation method:

\begin{yamlcode}[title=config/custom\_system.yml]
_include_yml: orthrus

featurization:
  used_method: fasttext
  fasttext:
    min_count: 2
    alpha: 0.01
    window_size: 3
    negative: 3
    num_workers: 1
batching:
  intra_graph_batching:
    used_methods: edges
training:
  encoder:
    used_methods: sage
    sage:
      activation: relu
      num_layers: 2
evaluation:
  node_evaluation:
    threshold_method: max_val_loss
    use_kmeans: False
\end{yamlcode}

\texttt{\_include\_yml} specifies the base configuration; unspecified parameters are inherited. This reduces configuration overhead and keeps variants aligned with the base system's settings.

The variant can be executed like any other system:

\begin{bashcode}[title=Run Custom System Config]
./run.sh custom_system CADETS_E3
\end{bashcode}

This workflow supports systematic exploration of the design space (\eg swapping featurization, encoders, decoders, or thresholding), enabling the ablation studies recommended by Arp et al.~\cite{arp2022and}.
Alternatively, any YAML parameter can be overridden from the CLI using dot notation:

\begin{bashcode}[title=Custom Experiment from CLI]
./run.sh orthrus CADETS_E3 \
	--featurization.used_method=fasttext \
	--training.encoder.used_methods=sage \
	[...]
\end{bashcode}

\subsection{Hyperparameter Tuning}

Fair evaluation requires consistent hyperparameter tuning across all benchmarked systems~\cite{arp2022and, bilot2025simpler}. \system streamlines tuning by combining its caching pipeline a specifc type of configueration YML file to describe the hyperparameter space to search.
For example, to grid-search over learning rate and hidden dimension:

\begin{yamlcode}[title=tuning\_custom\_system.yml]
method: grid

parameters:
  training.lr:
    values: [0.001, 0.0001]
  training.node_hid_dim:
    values: [32, 64, 128, 256]
\end{yamlcode}

We can the launch the sweep by enabling tuning mode:

\begin{bashcode}[title=Launch Hyperparameter Sweep]
./run.sh custom_system CADETS_E3 \
	--tuning_mode=hyperparameters \
	--exp=my_sweep_name \
	--project=my_project
\end{bashcode}

\system iteratively runs for each combination of hyperparameters (here for a total of 8 runs), loads the sweep configuration, overrides the corresponding parameters in the base system, and logs all runs to W\&B for analysis. 
After selecting the best settings in W\&B, users can store them in a tuned configuration that overrides the default system hyperparameters.
The tuned system can be run with:

\begin{bashcode}[title=Run Tuned System]
./run.sh custom_system CADETS_E3 --tuned
\end{bashcode}

To support fair comparison, we recommend applying the same tuning protocol to all systems (including baselines); \system provides pre-defined sweep configurations for each supported system.

\subsection{Ablation Studies}

Ablation studies isolate the contribution of individual components~\cite{arp2022and}. \system supports ablations using the same sweep mechanism by enumerating component choices. For example:

\begin{yamlcode}[title=ablation\_custom\_system.yml]
method: grid

parameters:
  featurization.used_method:
    values: [fasttext, word2vec]
  training.encoder.used_methods:
    values: [sage, graph_attention, none]
\end{yamlcode}

We can launch the ablation sweep by specifying the configuration file:

\begin{bashcode}[title=Launch Ablation Sweep]
./run.sh custom_system CADETS_E3 \
	--tuning_mode=hyperparameters \
	--tuning_file_path=systems/default/ablation_custom_system.yml
\end{bashcode}

All hyperparameter combinations are executed and logged to Weights \& Biases (W\&B) as part of a sweep~\cite{sweeps}, which is a specialized run type designed for systematic hyperparameter exploration.

\noindent\textbf{Parallelized Hyperparameter Tuning.}
To launch multiple concurrent runs, first retrieve the sweep ID from the W\&B interface, then execute the same command multiple times while specifying \texttt{--sweep\_id=PROJECT/ID} with the corresponding sweep identifier. Each run automatically evaluates the next available hyperparameter configuration in the sweep.
When performing concurrent hyperparameter tuning, it is important to monitor GPU memory usage and overall CPU load, as resource saturation may lead to memory overflows or significant runtime slowdowns.
For example, on a machine equipped with a 3.2 GHz 64-core CPU, 1 TB of system memory, and an 80 GB NVIDIA GPU, we observed that up to four concurrent \textsc{Orthrus} runs could be executed without noticeable per-run overhead or GPU memory exhaustion.

\subsection{Batching and Sampling}

Provenance graphs can reach millions of nodes and edges. \system provides flexible batching and sampling strategies to scale training while balancing memory usage, throughput, and learning effectiveness.

\input{fig/batching}

\noindgras{Batching Strategies.}
Graphs are constructed in the first stage with a default temporal granularity of 15 minutes. To refine granularity without rebuilding from construction, \system provides three batching strategies (Figure~\ref{fig:batching}) that can be configured via YAML and applied sequentially:

\begin{itemize}[leftmargin=*]
	\item \textbf{Global batching} Partition a flattened graph spanning the dataset into subgraphs using criteria such as number of edges or time duration (minutes).
	
	\item \textbf{Intra-graph batching} Further partition each constructed graph, enabling finer-grained temporal segmentation.
	
	\item \textbf{Inter-graph batching} Stack multiple graphs into a mini-batch without overlap, following PyTorch Geometric's batching semantics~\cite{pyg}.
\end{itemize}

Batch size trades off throughput and context against memory and model limitations. Larger batches improve GPU utilization and capture longer temporal context but increase memory use and may exacerbate over-squashing via high in-degree. Smaller batches reduce memory pressure and enable finer-grained aggregation but slow training and may miss longer-range patterns. Time-based batching preserves temporal dynamics but yields variable graph sizes; edge-based batching yields more uniform sizes but does not explicitly encode time.

\noindgras{TGN Last Neighbor Sampling.}
For systems based on Temporal Graph Networks (TGN)~\cite{rossi2020temporal}, such as \textsc{Kairos} and \textsc{Orthrus}, \system provides TGN-specific sampling support. In TGN, predictions for batch time $t$ condition on each node's recent neighbors observed in prior batches. 
\system pre-computes the required attributes during preprocessing and exposes them to the TGN encoder during training.

\subsection{Instability Measurement}
\label{sec:instability}

Neural network predictions can vary across runs due to random initialization and stochastic training~\cite{bhojanapalli2021reproducibility}, yet results are often reported from a single run. Following~\cite{bilot2025simpler}, \system quantifies instability by running the pipeline multiple times and aggregating metrics:

\begin{bashcode}[title=Instability Measurement]
./run.sh orthrus CADETS_E3 \
	--experiment=run_n_times
\end{bashcode}

Parameters in \texttt{run\_n\_times.yml} control the number of runs (\texttt{iterations}, default: 5) and the stage to restart from (\texttt{restart\_from}, default: \texttt{feat\_training}). These can also be overridden via CLI; for example, to run 10 iterations starting from encoder--decoder training:

\begin{bashcode}[title=Custom Instability Configuration]
./run.sh orthrus CADETS_E3 \
	--experiment=run_n_times \
	--experiment.uncertainty.deep_ensemble.iterations=10 \
	--experiment.uncertainty.deep_ensemble.restart_from=gnn_training
\end{bashcode}

\system reports each metric as \texttt{*\_mean}, \texttt{*\_std}, and \texttt{*\_std\_rel} (relative standard deviation):

\begin{equation}
	\tilde{\sigma}_{\text{metric}} = \frac{\sigma_{\text{metric}}}{\overline{\text{metric}}} \times 100
\end{equation}

\texttt{*\_std\_rel} normalizes variability by the mean, enabling comparisons across systems with different absolute performance. High values indicate that single-run results may be unrepresentative.

\begin{figure}[t]
	\centering
	\includegraphics[width=\columnwidth]{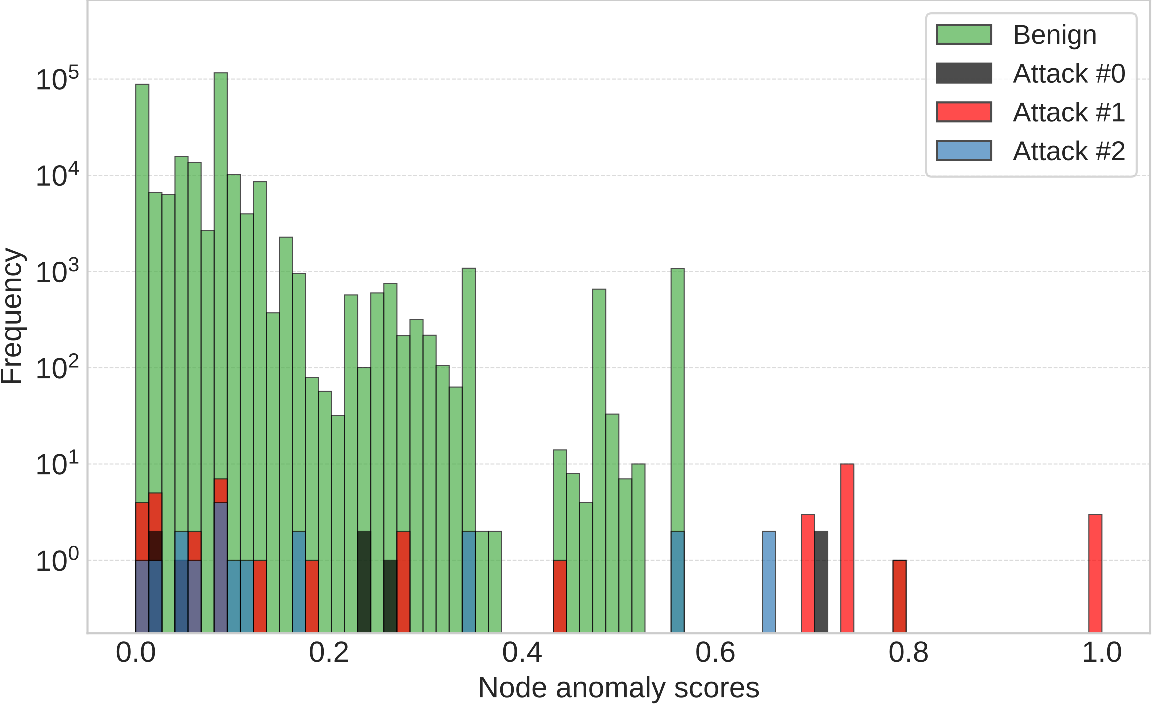}
	\caption{Distribution of (normalized) predicted anomaly scores for each attack in a dataset (\texttt{E3-CADETS} dataset).}
	\label{fig:distrib}
\end{figure}

\subsection{Metrics and Visualization}
\label{sec:viz}

\system runs inference on the test set using model checkpoints from each epoch and computes over 30 metrics. These include post-threshold metrics (precision, recall, F1), pre-threshold metrics (average precision, AUC-ROC, ADP~\cite{bilot2025simpler}), and PIDS-specific metrics such as the area under the discrimination curve (Figure~\ref{fig:rank}), which measures how consistently attack entities receive higher anomaly scores. All metrics are logged to W\&B and historized for epoch-level analysis.

\system also generates visualizations to aid interpretation. Score distributions show separation between benign and attack activity (Figure~\ref{fig:distrib}), while rank-based plots highlight discrimination among top-ranked entities (Figure~\ref{fig:rank}). W\&B integration additionally provides real-time monitoring of resource usage (CPU/GPU utilization and memory), storage, and network activity.

%% file: fig/batching.tex
\begin{figure}[t]
\centering
\resizebox{\columnwidth}{!}{%
\begin{tikzpicture}[
    scale=1.1,
    every node/.style={transform shape},
    graph/.style={
        rectangle,
        draw=black,
        minimum width=0.7cm,
        minimum height=0.7cm,
        align=center,
        font=\tiny,
        rounded corners=2pt
    },
    smallgraph/.style={
        rectangle,
        draw=black,
        minimum width=0.6cm,
        minimum height=0.6cm,
        align=center,
        font=\tiny,
        rounded corners=2pt
    },
    mergedgraph/.style={
        rectangle,
        draw=black,
        minimum width=1.9cm,
        minimum height=0.7cm,
        align=center,
        font=\tiny,
        rounded corners=2pt,
        fill=blue!15
    },
    batchbox/.style={
        rectangle,
        draw=black!40,
        fill=gray!20,
        line width=1.5pt,
        rounded corners=4pt,
        inner sep=4pt
    },
    arrow/.style={->, >=Stealth, semithick, black},
    timelabel/.style={font=\tiny, text=black!80},
    arrowlabel/.style={font=\tiny\itshape, text=black!70},
    descriptionlabel/.style={font=\tiny, text=black!80, align=center},
    sectiontitle/.style={font=\scriptsize\bfseries, anchor=west}
]

\node[sectiontitle] at (-2.9, 0.4) {Global batching};

\node[graph, fill=orange!25] (g1_glob) at (-2.1, -0.4) {$G_1$};
\node[graph, fill=blue!20] (g2_glob) at (-1.25, -0.4) {$G_2$};
\node[graph, fill=magenta!20] (g3_glob) at (-0.4, -0.4) {$G_3$};

\node[timelabel] at (-2.1, -0.9) {15min};
\node[timelabel] at (-1.25, -0.9) {15min};
\node[timelabel] at (-0.4, -0.9) {15min};

\draw[arrow] (0.1, -0.4) -- (0.6, -0.4);
\node[arrowlabel, above] at (0.35, -0.3) {flatten};

\node[mergedgraph] (merged) at (1.6, -0.4) {\tiny $G_1$+$G_2$+$G_3$};

\draw[arrow] (2.6, -0.4) -- (3.1, -0.4);
\node[arrowlabel, above, align=center] at (2.85, -0.15) {10-min\\[-1pt]partition};

\node[smallgraph, fill=gray!35] (g1p) at (3.6, -0.4) {$G'_1$};
\node[smallgraph, fill=gray!35] (g2p) at (4.25, -0.4) {$G'_2$};
\node[font=\small] at (4.7, -0.4) {...};
\node[smallgraph, fill=gray!35] (gnp) at (5.2, -0.4) {$G'_n$};

\node[timelabel] at (3.6, -0.9) {10min};
\node[timelabel] at (4.25, -0.9) {10min};
\node[timelabel] at (5.2, -0.9) {10min};

\node[descriptionlabel] at (4.4, -1.3) {Each graph has a fixed size\\(e.g. edges or minutes)};

\node[sectiontitle] at (-2.9, -1.85) {Intra-graph batching};

\node[graph, fill=orange!25] (g1_intra) at (-2.1, -2.6) {$G_1$};
\node[graph, fill=blue!20] (g2_intra) at (-1.25, -2.6) {$G_2$};
\node[graph, fill=magenta!20] (g3_intra) at (-0.4, -2.6) {$G_3$};

\node[timelabel] at (-2.1, -3.1) {15min};
\node[timelabel] at (-1.25, -3.1) {15min};
\node[timelabel] at (-0.4, -3.1) {15min};

\draw[arrow] (0.1, -2.6) -- (0.6, -2.6);
\node[arrowlabel, above, align=center] at (0.35, -2.35) {10-min\\[-1pt]partition};

\node[smallgraph, fill=orange!25] at (1.1, -2.6) {$G'_1$};
\node[smallgraph, fill=orange!15] at (1.75, -2.6) {$G'_2$};
\node[smallgraph, fill=blue!20] at (2.5, -2.6) {$G'_3$};
\node[smallgraph, fill=blue!10] at (3.15, -2.6) {$G'_4$};
\node[smallgraph, fill=magenta!20] at (3.9, -2.6) {$G'_5$};
\node[smallgraph, fill=magenta!12] at (4.55, -2.6) {$G'_6$};

\node[timelabel] at (1.1, -3.1) {10min};
\node[timelabel] at (1.75, -3.1) {5min};
\node[timelabel] at (2.5, -3.1) {10min};
\node[timelabel] at (3.15, -3.1) {5min};
\node[timelabel] at (3.9, -3.1) {10min};
\node[timelabel] at (4.55, -3.1) {5min};

\node[descriptionlabel] at (2.8, -3.55) {Same approach, but partitions\\within each graph};

\node[sectiontitle] at (-2.9, -4.15) {Inter-graph batching};

\node[graph, fill=orange!25] (g1_inter) at (-2.1, -4.9) {$G_1$};
\node[graph, fill=blue!20] (g2_inter) at (-1.25, -4.9) {$G_2$};
\node[graph, fill=magenta!20] (g3_inter) at (-0.4, -4.9) {$G_3$};

\node[timelabel] at (-2.1, -5.4) {15min};
\node[timelabel] at (-1.25, -5.4) {15min};
\node[timelabel] at (-0.4, -5.4) {15min};

\draw[arrow] (0.2, -4.9) -- (0.7, -4.9);
\node[arrowlabel, above] at (0.35, -4.65) {batch of 3 graphs};

\begin{scope}[shift={(2.15, -4.9)}]
    \node[batchbox, minimum width=2.1cm, minimum height=1.1cm] at (0, 0) {};
    \node[graph, fill=orange!25, minimum width=0.55cm, minimum height=0.55cm] at (-0.55, 0) {$G_1$};
    \node[graph, fill=blue!20, minimum width=0.55cm, minimum height=0.55cm] at (0, 0) {$G_2$};
    \node[graph, fill=magenta!20, minimum width=0.55cm, minimum height=0.55cm] at (0.55, 0) {$G_3$};
\end{scope}

\node[descriptionlabel] at (1.85, -5.8) {Standard mini-batching.\\Multiple graphs are merged into a large one};

\end{tikzpicture}%
}
\caption{Batching strategies in \system. \textbf{Global}: flatten all graphs and repartition into fixed-size chunks. \textbf{Intra-graph}: partition within each graph. \textbf{Inter-graph}: group multiple graphs into a mini-batch.}
\label{fig:batching}
\end{figure}

%% file: sections/rw.tex
\noindgras{PIDS surveys and systematizations.} Inam \etal~\cite{inam2023sok} provide a comprehensive systematization of provenance-based intrusion detection and investigation. Bilot \etal~\cite{bilot2023graph} survey the broader use of GNNs for intrusion detection. These works categorize existing techniques but do not provide implementation or evaluation frameworks.

\noindgras{Reproducibility studies.} Abrar \etal~\cite{abrar2025rep} studied the reproducibility of deep learning-based PIDSs and found that published results are often difficult to reproduce due to missing artifacts and insufficient implementation detail. Bilot \etal~\cite{bilot2025simpler} evaluated eight systems and identified methodological shortcomings, including prediction instability and data snooping. \system addresses these issues through standardized preprocessing, consistent evaluation, and built-in experimental utilities.

\noindgras{Benchmark frameworks.} ProvMark~\cite{chan2019provmark} benchmarks the expressiveness of provenance capture systems, and Xanthus~\cite{han2020xanthus} orchestrates provenance data collection. Both emphasize capture rather than detection system development and evaluation. To our knowledge, \system is the first framework purpose-built for PIDS development and evaluation.

\noindgras{ML security evaluation guidelines.} Arp \etal~\cite{arp2022and} provide evaluation guidelines for machine learning in security, highlighting pitfalls such as data snooping, weak baselines, and missing ablations. \system operationalizes these recommendations via systematic hyperparameter tuning, ablation support, and multi-run instability measurement.

%% file: sections/discussion.tex
Consolidating and evaluating multiple PIDSs within a unified framework surfaced several open challenges that merit further study:

\noindgras{Prediction instability.} Instability remains a central limitation of current self-supervised architectures. 
The training objective (predicting properties of provenance graphs) is misaligned with the evaluation objective (detecting attacks), so it is often unclear which learned features drive detection. 
As a result, some runs learn useful representations while others do not. While the root cause remains open, repeating experiments (Section~\ref{sec:instability}) yields more reliable performance estimates. 
Future work could explore weakly supervised or few-shot approaches that incorporate limited attack supervision during training.

\noindgras{Lack of benign anomaly benchmarks.} Existing benchmark datasets typically treat the world as binary (benign vs.\ attack) and rarely label \emph{benign anomalies} (legitimate but uncommon behavior) as a separate class. 
As a result, self-supervised PIDSs trained on ``normal'' traces often assign high anomaly scores to rare yet acceptable events (\eg software updates, administrative maintenance, or bursty but authorized activity), which can inflate false positives. 
Further, real deployments face shifting definitions of normal across hosts, workloads, and time, and some datasets already contain unlabeled benign irregularities that confound evaluation. 
Benchmarks that explicitly annotate benign anomalies (or provide workload/maintenance labels) would enable more realistic measurement of detection specificity and help disentangle ``rare but benign'' from truly malicious behavior.

\noindgras{Robustness evaluation.} Concept drift and adversarial manipulation are practical threats to deployed PIDSs. 
We plan to extend \system with concept-drift detection and adaptation~\cite{jordaney2017transcend, yang2021cade}, and with adversarial robustness evaluation~\cite{Goyal2023SometimesYA, Mukherjee2023EvadingPM}.

\noindgras{Scope of supported systems.} \system currently focuses on self-supervised PIDSs.
Extending support to supervised and rule-based approaches~\cite{wang2024incorporating, milajerdi2019holmes} would enable broader comparisons and is planned for future releases.

%% file: sections/community.tex
We hope \system can provide a shared evaluation baseline for PIDS research. 
Contributions are welcome: including implementations of additional systems, new datasets, and improved components.
The modular architecture and YAML-based configuration are intended to make extending the framework straightforward and low-overhead.

%% file: sections/conclusion.tex
We presented \system, a unified framework for developing and evaluating provenance-based intrusion detection systems. By consolidating eight state-of-the-art PIDSs into a modular architecture with standardized preprocessing, consistent ground-truth labels, and integrated experimental utilities, \system mitigates key reproducibility and engineering costs that slow progress in this area.

\system enables rapid prototyping via component reuse, supports systematic ablations and hyperparameter tuning, quantifies run-to-run prediction instability, and facilitates fair comparisons under consistent evaluation protocols. We view common evaluation standards as essential for sustained progress, and hope \system helps move the community toward that goal. We welcome adoption and contributions that extend and improve the framework.

%% file: sections/availability.tex
The code and documentation of the framework will be made publicly available upon acceptance of the paper.

%% file: sections/ethics.tex
We have considered the ethical implications of our work~\cite{kohno2023ethical}.
\system aims to improve the rigor and reproducibility of provenance-based security research. 
All datasets we use are publicly available and, to the best of our knowledge, contain no sensitive personal information. 
\system is an evaluation and development framework and does not introduce new attack capabilities.